\documentclass[prb,aps,super,
twocolumn,
reprint,
groupedaddress,superscriptaddress,
amsfonts,amssymb,amsmath,floatfix,
showkeys,showpacs,a4paper]{revtex4-1}

\usepackage{graphicx}
\usepackage[english]{babel}
\usepackage{hyperref}
\usepackage{upgreek}
\usepackage{natbib}
\usepackage{lineno}
\usepackage{color}
\usepackage{lipsum}

\begin{document}

\title{
High entropy oxides: An emerging prospect for magnetic rare earth - transition metal perovskites
}

\author{Ralf Witte}\email{ralf.witte@kit.edu}
\affiliation{Institute of Nanotechnology, Karlsruhe Institute of Technology, 
 76344 Eggenstein-Leopoldshafen, Germany}
 \author{Abhishek Sarkar}
\affiliation{Institute of Nanotechnology, Karlsruhe Institute of Technology, 76344 Eggenstein-Leopoldshafen, Germany}
\affiliation{KIT-TUD-Joint Research Laboratory Nanomaterials, Technical University Darmstadt,  64287 Darmstadt, Germany}
\author{Robert Kruk}
\affiliation{Institute of Nanotechnology, Karlsruhe Institute of Technology, 
 76344 Eggenstein-Leopoldshafen, Germany}
 \author{Benedikt Eggert}
 \affiliation{Faculty of Physics and Center for Nanointegration Duisburg-Essen (CENIDE), University of Duisburg-Essen, Lotharstr. 1, 47048 Duisburg, Germany}
\author{Richard A. Brand}
\affiliation{Institute of Nanotechnology, Karlsruhe Institute of Technology, 
 76344 Eggenstein-Leopoldshafen, Germany}
\affiliation{Faculty of Physics and Center for Nanointegration Duisburg-Essen (CENIDE), University of Duisburg-Essen, Lotharstr. 1, 47048 Duisburg, Germany}
\author{Heiko Wende}
 \affiliation{Faculty of Physics and Center for Nanointegration Duisburg-Essen (CENIDE), University of Duisburg-Essen, Lotharstr. 1, 47048 Duisburg, Germany}
\author{Horst Hahn}
\affiliation{Institute of Nanotechnology, Karlsruhe Institute of Technology, 
 76344 Eggenstein-Leopoldshafen, Germany}
\affiliation{KIT-TUD-Joint Research Laboratory Nanomaterials, Technical University Darmstadt,  64287 Darmstadt, Germany}

\date{\today}
\begin{abstract}

It has been shown that  oxide ceramics containing multiple transition and/or rare-earth elements in equimolar ratios have a strong tendency to crystallize in simple single phase structures, stabilized by the high configurational entropy. In analogy to the metallic alloy systems, these oxides are denoted high entropy oxides (HEOs). The HEO concept allows to access hitherto uncharted areas in the multi-element phase diagram. Among the already realized structures there is the highly complex class of rare earth - transition element perovskites. 
This fascinating class of materials generated by applying the innovative concept of high entropy stabilization provides a new and manyfold  research space with promise of  discoveries of unprecedented properties and phenomena. 
The present study provides a first investigation of the  magnetic properties of  selected compounds of this novel class of materials. Comprehensive studies by DC and AC magnetometry are combined  with element specific spectroscopy in order to understand the interplay between  magnetic exchange and the high degree of chemical disorder in the systems. We observe a predominant antiferromagnetic behavior in the single phase materials, combined with a small ferromagnetic contribution. The latter can be attributed to either  
small ferromagnetic clusters or configurations in the antiferromagnetic matrix or a possible spin canting.  
In the long term perspective it is proposed to screen the properties of this family of compounds with high throughput methods, including combined experimental and theoretical approaches.

\end{abstract}

\pacs{}

\maketitle

\section{Introduction}

High entropy oxides (HEOs) represent a new class of oxide systems that have already attracted significant research interest since their recent discovery \cite{Rost2015}. The key point of the high entropy stabilization concept  is the combination of a large number of cations (usually five or more) in solid solution  in equiatomic proportions, which often results in the formation of a single phase structure, overcoming the usual enthalpy driven phase separation usually encountered in heavily doped systems\cite{Cheng2005c}.
In this  way, single phase compounds with compositions  in the center of a complex phase diagram can be produced, which are  seldom studied. Such compounds, stabilized by configurational entropy, will be increasingly stable with increasing temperature.
Several compositions and elemental combinations each resulting in different crystal structures, such as rocksalt, fluorite, spinels and perovskite, have been stabilized using the HEO concept\cite{Rost2015,Sarkar2017b,Dabrowa2018,Jiang2018,Sharma2018}. 

In many of the studied cases (such as rocksalt\cite{Rost2015,Berardan2016a}, flourite\cite{Chen2018} and perovskite type HEOs\cite{Sarkar2017a}) it is well understood that the large configurational entropy of the systems dominates the Gibbs free energy of formation and eventually compensates any positive enthalpic contributions.   
 The configurational entropy of a system increases with the number of  different elements distributed over the cation lattice site and it attains a maximum when all the constituent elements are present in equiatomic amounts\cite{Rost2015}. Apart from the interesting structural  ramifications, this distinct design concept may also allow for the fine adjustment of the functional properties. Some examples of already reported tailorable properties in the HEOs are high room temperature Li$^{+}$ conductivity\cite{Berardan2016}, catalytic properties\cite{Chen2018b}, colossal dielectric constants\cite{Berardan2016a},  superior capacity retention capabilities\cite{Sarkar2018b}, narrow  and adjustable band gaps\cite{Sarkar2017b}, to name a few. However, as the field of HEOs is at its early stage many of the  material characteristics still wait to be investigated. One of such yet unexplored fields is the magnetism of HEOs, where so far only a single study exists. This study however focuses on the magnetic  interaction of a rocksalt HEO with a magnetic layer in a thin film heterostructure, rather than on the intrinsic magnetic properties of the HEO compound \cite{Meisenheimer2017}.

\begin{figure}[htb]
\[\includegraphics[width=1\columnwidth]{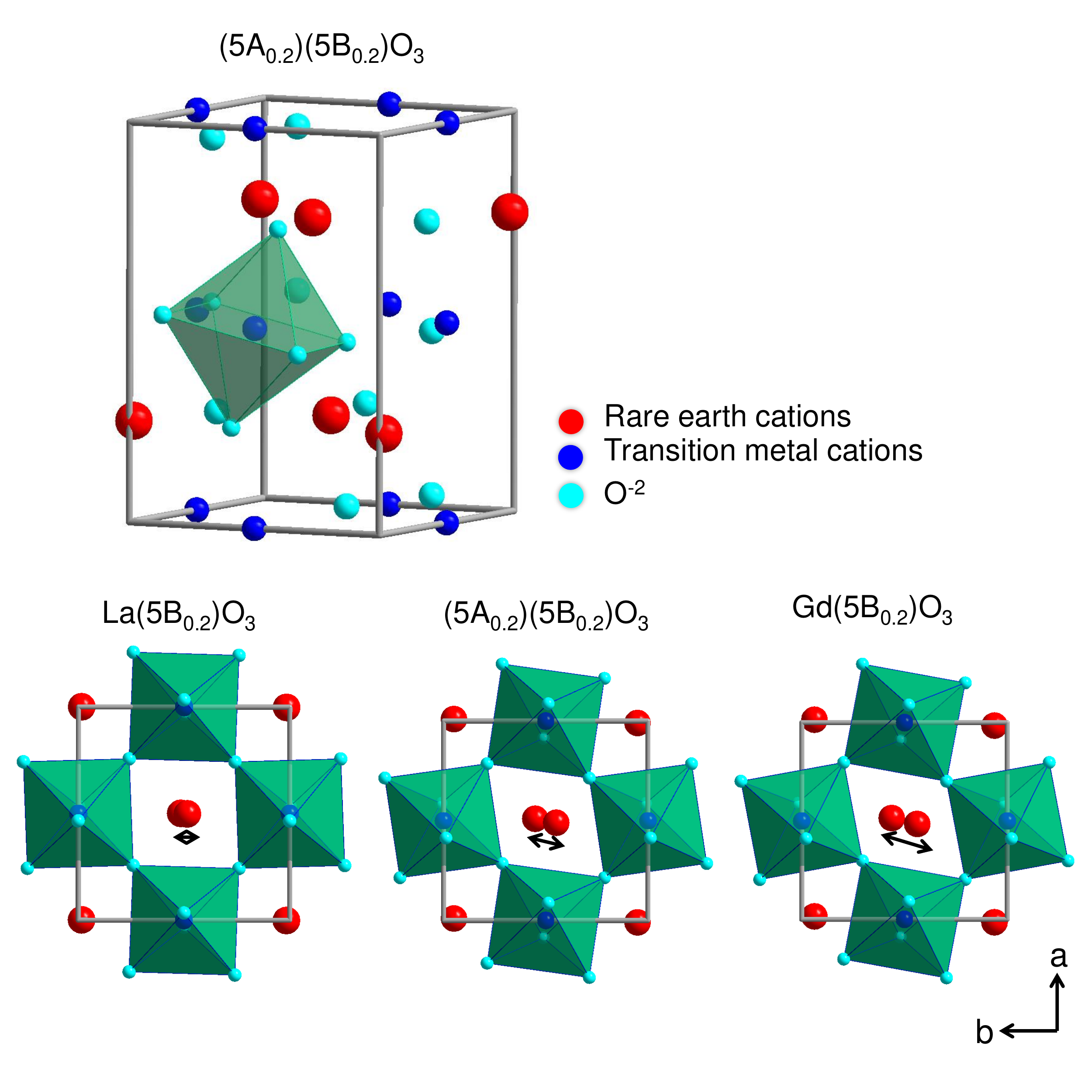}\]
\caption{Upper part: Crystal structure of a representative orthorhombic ($Pbnm$) PE-HEO, (5A$_{0.2}$)(5B$_{0.2})$O$_3$. Lower part: Illustration of the increasing magnitude of tilting of the BO$_6$ polyhedra observed in PE-HEOs along [001] axis with decreasing tolerance factor (larger deviation from a an ideal lattice with tolerance factor$=$1) for: La(5B$_{0.2})$O$_3$, (5A$_{0.2}$)(5B$_{0.2})$O$_3$, and Gd(5B$_{0.2})$O$_3$.}
\label{fig_structure}
\end{figure}

Comparing the different HEO parent oxides structures, one can see that perovskites, with the general formula ABO$_3$, form one of the most  complex and recognized class of oxide materials, see Fig.\ref{fig_structure} for illustration.  
Amongst perovskites, rare earth and transition metal based oxides have been by far the most extensively studied systems over the last few decades due to their unique properties, from both a fundamental as well as an application point of view.  Here, A represents any number of different rare earth ions (RE), and B any number of different transition metal ions (TM).
 Mixing REs on the A site and/or TMs on the B site  allows for adjustment of structural and therefore possibly also functional properties over a wide range. The rich choice of promising characteristics and complex physics found in the parent perovskite compounds, such as e.g., multiferroic effects\cite{Cheong2007}, catalytic activity\cite{Mawdsley2008}, electronic\cite{Goodenough2001}, electrochemical and related transport properties\cite{Koep2006,Skinner2003},  make them promising candidates for a broad range of engineering applications.
 
 Many of the interesting properties in the perovskites in general are directly related to their crystal structure. Crystal structure stability in perovskites is largely governed by the Goldschmidt's tolerance factor, which is a function of the constituents ionic radii\cite{Tilley2016}. Hence, tailoring the properties for desired applications often starts by altering the cationic radii,  realized either by doping or substitution of specific cations. Fig.\,\ref{fig_structure} showcases the different degree of octahedral tilting for three representative compounds with different tolerance factor. 

However, doping or substitutional approaches very often have the limitation in a sense that only relatively small levels of doping can be achieved, due to either the presence of phase boundaries leading to a different structure with undesired properties or phase segregation according to the equilibrium thermodynamics\cite{McBride2016,Jiang2008}.

In this study, the magnetic properties of RE - TM based high entropy perovskites are investigated. Magnetic properties of conventional RE - TM perovskites (ABO$_3$) have been a major research interest for almost half a century\cite{Rao1996,Eibschutz1967a}. The interlink between their magnetic properties and crystal structure has been extremely important as any type of structural changes, such as lattice distortion or tilting of the BO$_6$ octahedra (see Fig.\,\ref{fig_structure}), have often shown a decisive impact on material properties. In this case study the observed unique magnetic properties of the perovskite based high entropy oxides (PE-HEO) mainly originate from the presence of multinary TM cations, as they govern magnetic exchange at finite temperatures. A combination of careful magnetometry and M\"ossbauer spectroscopy experiments has been used to unravel their complex magnetic behavior,  dominated by competing antiferromagnetic (AFM) and ferromagnetic (FM) interactions in the TM sublattice. Despite the large number of constituent ions the effect of cationic radii, as measured by changes in the Goldschmidt's tolerance factor, on the magnetic ordering temperature of the compounds has also been observed.

\section{Experimental details}\label{sec:exp}

\subsection{Synthesis and structural characterization}
\label{sec:exp_syn}
PE-HEOs were synthesized by using the nebulized spray pyrolysis (NSP) method. This is an aerosol based synthesis technique in which the decomposition of the precursor solution at elevated temperature leads to the formation of the desired phase \cite{Djenadic2014}. This phase can either be the final product, or in some cases an intermediate phase which is then given further heat treatments. In this study the aqueous precursor solutions are mixtures in the appropriate concentrations  of the corresponding nitrate salts of the constituent cations. The temperature of the hot wall reactor was maintained at 1050\,$^{\circ}$C during the synthesis. The as-synthesized powders  were additionally heat treated at 1200\,$^{\circ}$C for 2\,hours in air, in order to achieve the final single PE-HEO phase. The detailed description of the synthesis procedure  is reported elsewhere \cite{Sarkar2017a}. 
Five of the systems investigated, have a single RE A-site cation and mixed (TM) B-site cations, namely, 
\begin{itemize}
\item[]Gd(Co$_{0.2}$Cr$_{0.2}$Fe$_{0.2}$Mn$_{0.2}$Ni$_{0.2}$)O$_3$, \item[]La(Co$_{0.2}$Cr$_{0.2}$Fe$_{0.2}$Mn$_{0.2}$Ni$_{0.2}$)O$_3$, \item[]Nd(Co$_{0.2}$Cr$_{0.2}$Fe$_{0.2}$Mn$_{0.2}$Ni$_{0.2}$)O$_3$, \item[]Sm(Co$_{0.2}$Cr$_{0.2}$Fe$_{0.2}$Mn$_{0.2}$Ni$_{0.2}$)O$_3$, \item[]Y(Co$_{0.2}$Cr$_{0.2}$Fe$_{0.2}$Mn$_{0.2}$Ni$_{0.2}$)O$_3$.
\end{itemize}

In addition, a 10 equiatomic cationic system,  
\begin{itemize}
\item[](Gd$_{0.2}$La$_{0.2}$Nd$_{0.2}$Sm$_{0.2}$Y$_{0.2}$)\\(Co$_{0.2}$Cr$_{0.2}$Fe$_{0.2}$Mn$_{0.2}$Ni$_{0.2}$)O$_3$ 
\end{itemize}
which features also a mixed RE A-site was also studied. 
In the  sections below, we will denote the  mixed A-site (Gd$_{0.2}$La$_{0.2}$Nd$_{0.2}$Sm$_{0.2}$Y$_{0.2}$) by (5A$_{0.2}$)  and use (5B$_{0.2}$) for the mixed B-site (Co$_{0.2}$Cr$_{0.2}$Fe$_{0.2}$Mn$_{0.2}$Ni$_{0.2}$).

Powder X-ray diffraction (XRD) patterns were recorded using a Bruker D8 diffractometer with Bragg-Brentano geometry using Cu-K$\alpha$ radiation with a Ni filter. Rietveld analysis of the XRD patterns, performed using TOPAS 5 refinement software,  confirmed that 4 out of 6 systems studied, crystallize into a single phase, pure orthorhombic ($Pbnm$) structure, which includes the chemically complex decanary system\cite{Sarkar2017a}. Transmission electron microscopy studies  evidence a homogeneous distribution of the multiple elements\cite{Sarkar2017a}. Two systems, Sm(5B$_{0.2}$)O$_3$ and Y(5B$_{0.2}$)O$_3$, show in addition small amounts  of non-perovskite type secondary phases (1.7\,wt.\% Sm$_2$O$_3$ and 3.2\,wt.\% Y$_2$O$_3$, 2.1\,wt.\% NiO respectively). Both of these minority phases have no implications for the analysis of the magnetic properties presented in the following, as they either are paramagnetic in the entire temperature regime or have an AFM transition temperature above the temperature regime  investigated here. Structural details of all these systems are tabulated in the Supplementary Information Table S1.

\subsection{Magnetic and M\"ossbauer characterization}

Magnetic characterization was performed using a Quantum Design MPMS3 Superconducting Quantum Interference Device (SQUID) vibrating sample magnetometer (VSM). After  the sample mass was carefully determined the samples were mounted in the dedicated Quantum Design powder sample holders  and subsequent magnetization measurements were done in VSM mode. Temperature dependent measurements were performed following a zero-field cooled (ZFC) - field cooled (FC) protocol: The sample was cooled in zero magnetic field down to 2\,K. Then the external field $\upmu_0 H$ was applied and the magnetization then measured during warming up to 400\,K (ZFC branch). Subsequently, the magnetization was measured with the magnetic field applied from 400\,K to 2\,K (FC branch). Magnetic field dependent $M$($\upmu_0 H$) measurements were also performed after cooling in zero magnetic field. In addition, some measurements were performed after deliberately cooling in a magnetic field denoted $\upmu_0 H_{\mathrm{FC}}$, which is highlighted in the respective figures and text.

$^{57}$Fe M\"ossbauer spectroscopy (MS) was carried out employing a $^{57}$Co:Rh source. Samples were measured in transmission geometry using a triangular sweep of the velocity scale. In-field measurements were realized with the magnetic field parallel to the $\gamma$ radiation. As it is conventionally done, all center shifts are given relative to $\alpha$-Fe at room temperature.

\section{Results and discussion}\label{sec:results}

This section is divided into two parts: Section\,\ref{La_B_O3} presents  results of detailed DC and AC magnetometry measurements as well as M\"ossbauer measurements of the 
La(5B$_{0.2}$)O$_3$
compound. In this material, it is possible to study the  physics of the magnetic exchange interaction in the B site sublattice independently of the RE lattice, as La$^{3+}$ carries no magnetic moment. The magnetic exchange interactions in these oxide systems are generally governed by indirect interactions. The most common interaction present here is the  superexchange interaction \cite{Anderson1950}, which couples the spins of two neighboring TM ions via hybridization with the oxygen orbitals. This B-O-B coupling can be effectively AFM or FM, depending on the geometrical characteristics of the bond (90 or 180\,$^{\circ}$) and the electronic configuration of the two coupled TM ions as summarized in the Kannamori-Goodenough rules \cite{Goodenough1958,Kanamori1959}.

Although AFM ordering dominates for most of the B1$^{x+}$-O$^{2-}$-B2$^{y+}$ couples, with $x,y$ being their respective oxidation states, there exist also combinations where a FM interaction prevails, such as e.g. Fe$^{3+}$-O$^{2-}$-Cr$^{3+}$ or Ni$^{2+}$-O$^{2-}$-Mn$^{4+}$ \cite{Kanamori1959}. The occurrence of the latter couple is for example observed in the ternary oxide LaNi$_{1-x}$Mn$_x$O$_{3}$ leading to FM long range order\cite{Sanchez2002} and can be present also in the HEO compounds as a local charge compensation mechanism. We consider the compounds to be fully oxidized and stochiometric, due to the high temperature annealing in air. Additionally,  the occurrence of the double exchange mechanism, which is a delocalized kinetic exchange via the oxygen observed in multivalent manganites \cite{Goodenough1955} leads to FM order. The La(5B$_{0.2}$)O$_3$ PE-HEO provides thus the possibility to study these competing exchange interactions in detail.
Moreover  La(5B$_{0.2}$)O$_3$ has a Goldschmidt tolerance factor (see section\,\ref{sec:exp_syn}) which is closest to 1. Thus it is the least distorted crystal lattice and so it makes a natural  starting point for the compositional sample series discussed in the next section of the manuscript. 

Section\,\ref{sec:A_5B0,2_O3} presents and discusses the results of magnetization measurements of the entire series  A(Co$_{0.2}$Cr$_{0.2}$Fe$_{0.2}$Mn$_{0.2}$Ni$_{0.2}$)O$_3$  (A=Gd, La, Nd, Sm, Y, (5A$_{0.2}$)) hence including the decenary system denoted (5A$_{0.2}$)(5B$_{0.2}$)O$_3$  and puts the findings of Sec.\,\ref{La_B_O3} in the context of the structural series.

\subsection{La(Co$_{0.2}$Cr$_{0.2}$Fe$_{0.2}$Mn$_{0.2}$Ni$_{0.2}$)O$_3$}
\label{La_B_O3}

\begin{figure}[htb]
\[\includegraphics[width=0.9\columnwidth]{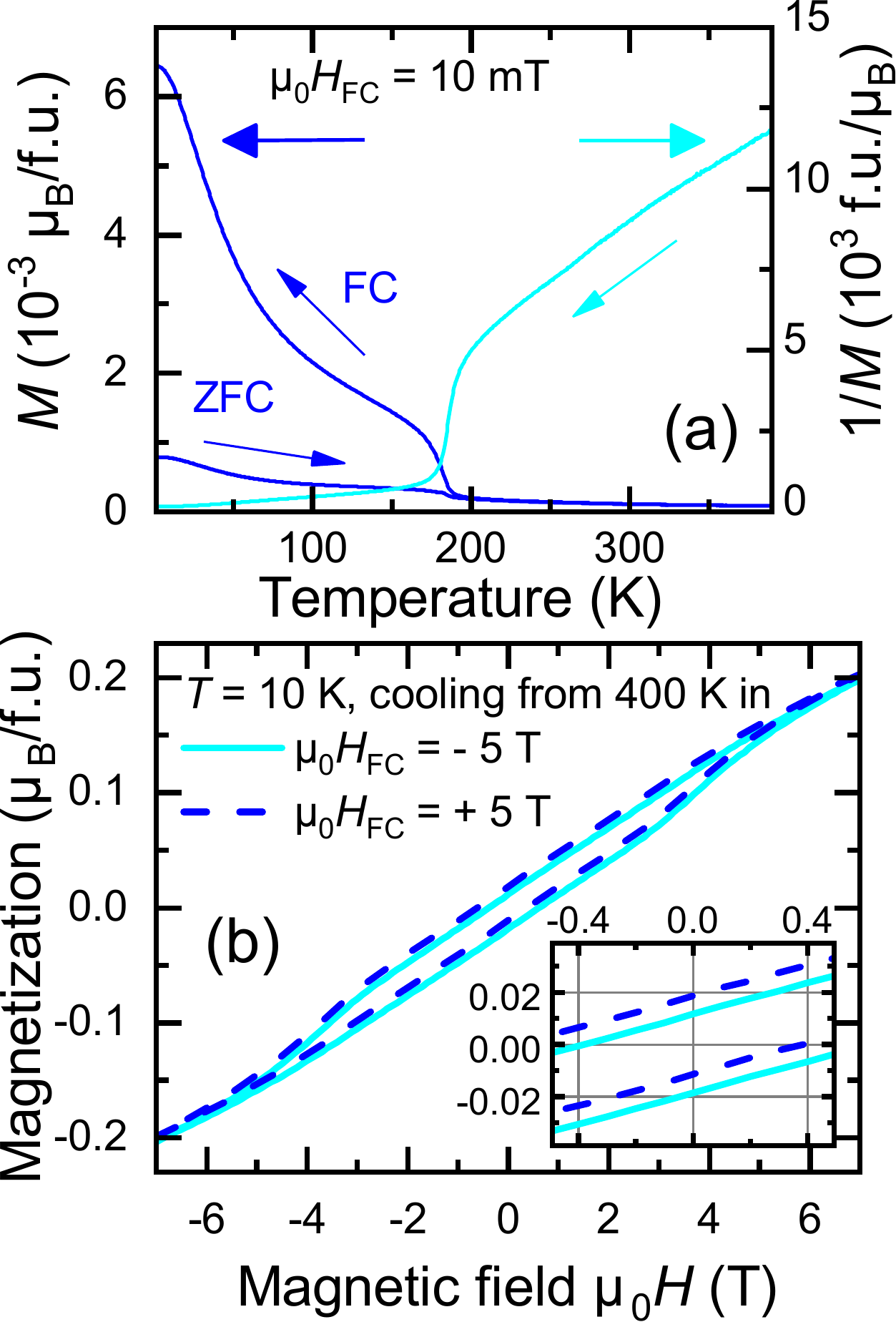}\]
\caption{(a) Temperature dependent magnetization after ZFC and in FC mode; the right-hand ordinate refers to the inverse magnetization. (b) Magnetization as function of the magnetic field $\upmu_{0}H$ of La(5B$_{0.2}$)O$_3$ at $T=10$\,K, after  FC in $\upmu_{0}H_{\mathrm{FC}}$=$\pm$5\,T. The inset shows the region around the center of the coordinate system, the same axis labels apply.}
\label{fig_m_ex_labo}
\end{figure}

Fig.\,\ref{fig_m_ex_labo} presents (a) temperature dependent magnetization $M$($T$)  (in $\upmu_0H=10$\,mT) and its inverse  as well as (b) magnetic field dependent measurements at 10\,K. The $M(T)$ behavior provides clear evidence for a magnetic phase transition at $T_{\mathrm{N}}=185$\,K, moreover the large differences between the ZFC and FC branches indicate the presence of large magnetic anisotropy.  
The inverse susceptibility also shows a clear magnetic transition and in addition a strong deviation from linear behavior at high temperatures above the transition. The latter is an indication of magnetic correlations existing even above the transition temperature prohibiting the extraction of the (average) effective paramagnetic moment $\mu_{\mathrm{eff}}$ from the linear part. The presence of these correlations is reasonable as the magnitude of the magnetic superexchange interactions of e.g., Fe-O-Fe\cite{Eibschutz1967a}, Cr-O-Cr\cite{Daniels2013,Prado-Gonjal2013} or Fe-O-Cr\cite{Ueda1998} couples is quite large.

Low temperature hysteresis measurements $M(\upmu_{0}H)$ after FC in $\upmu_{0}H_{\mathrm{FC}}$=$\pm$5\,T (Fig.\,\ref{fig_m_ex_labo}(b))  are nearly linear up to the highest attainable magnetic field of $\upmu_{0}H=7$\,T and show no sign of saturation. However, the presence of a  opening of the hysteresis  indicates that some magnetic moments or a projection of the  magnetic moment stays aligned at zero magnetic field (remnant magnetization).  Interestingly, this opening of the hysteresis curve extends even up to high magnetic field, resulting in a considerably large coercive field $\upmu_{0}H_{\mathrm{C}}$=3.6\,T of the ferromagnetic part of the curve at 10\,K, which is again a sign of strong magnetic anisotropies present in the sample. In fact, measuring the samples at $T<10$\,K results in so called minor loops, hence the accessible magnetic field is not sufficient to reverse the magnetization completely (see additional data in the supplementary material). This observation shows that the coercive field $\upmu_{0}H_{\mathrm{C}}$, which is already large at 10\,K,  is steeply increasing with further decreasing the temperature.

 \begin{figure}[htb]
\[\includegraphics[width=0.8\columnwidth]{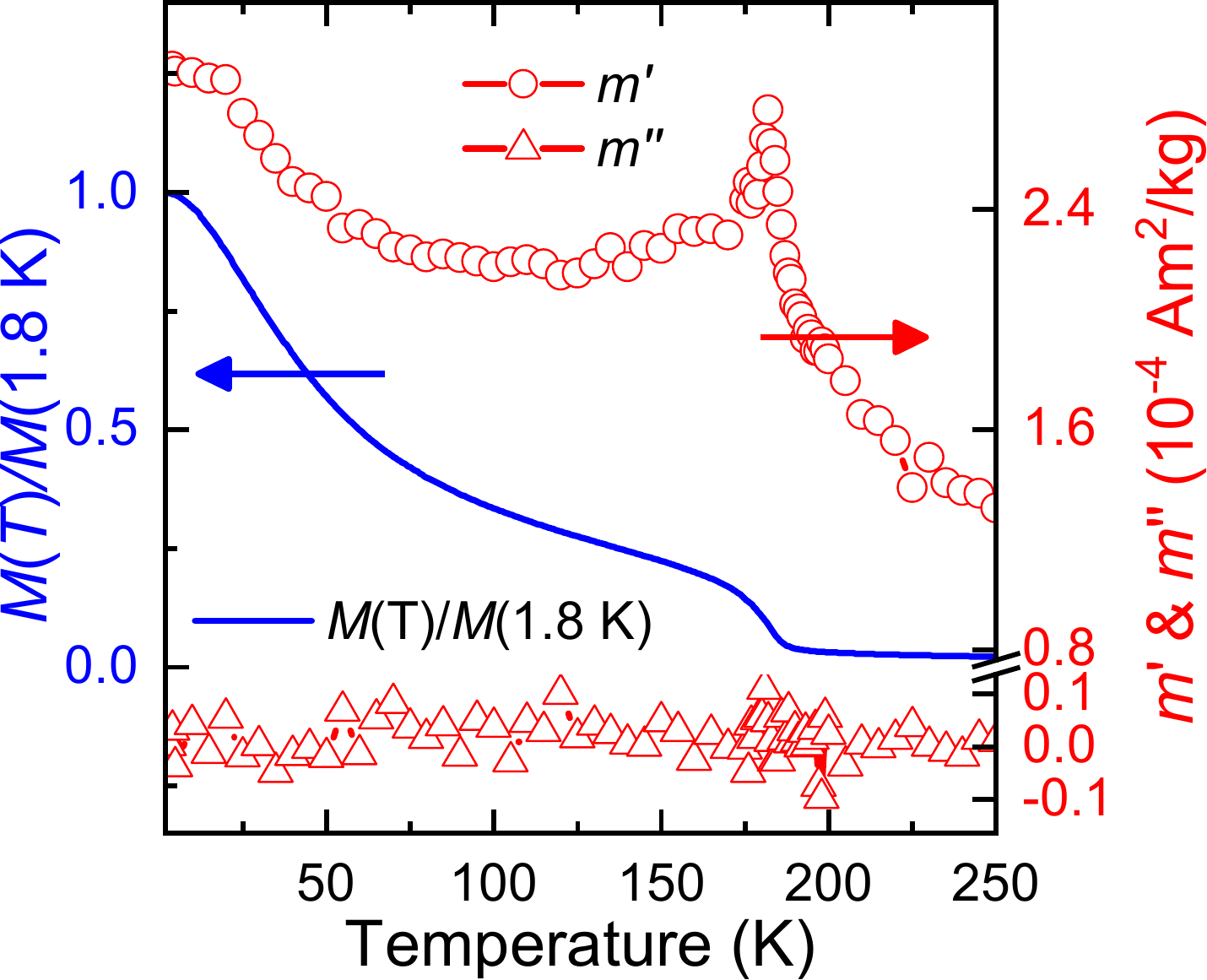}\]
\caption{In comparison to the AC magnetization $m'$ and $m''$, the normalized DC magnetization as function of temperature is shown.}
\label{fig_ex_labo}
\end{figure}

$M(\upmu_{0}H)$ curves measured after FC in $\pm$\,5\,T,  presented in Fig.\,\ref{fig_m_ex_labo}(b), show an obvious field offset from the center seemingly in both horizontal and vertical  directions. Analyzing the derivative of the entire curve (see supplementary material) shows that the FM part of the curve is not shifted along the field axis, as it would be expected for an exchange bias effect. However the vertical exchange bias (VEB) is real and amounts to a relative size of 25\%  with respect to the remnant magnetization.

The VEB is known to be a possible result of uncompensated spins in AFM materials \cite{Henne2016}, which align during FC giving a net magnetic moment, but which cannot be reoriented at low temperatures by the magnetic field due to their strong coupling to the AFM lattice. However in the PE-HEOs, we have additionally a precondition for competing FM and AFM exchange interactions, because e.g., the Fe-O-Cr or also mixed valence pairs are coupling ferromagnetically.   Assuming a simple binomial distribution one obtains a probability of about 10\% for finding more than e.g., three Cr$^{3+}$ ions as nearest neighbors of one Fe$^{3+}$ ion. These ensembles can act as small FM clusters in the AFM matrix, which get frozen and locked during FC. However the coupling to the surrounding AFM matrix results in a strong frustration and probably non-collinear arrangement of these magnetic moments at low temperatures. Moreover the exchange coupling with the AFM matrix  also explains the extraordinarily large coercive fields of the FM component at low temperatures (see supplementary materials, opening of the hysteresis up to nearly 7\,T).

 The VEB effect has been further studied as function of temperature from which the sample was cooled in the magnetic field, here denoted $T_{\mathrm{FC}}$.  
 The VEB is largest  when the magnetic field is applied above the magnetic transition, reaching a relative value of 25\%. Yet when the field is applied below the transition temperature one expect that the effect is directly vanishing, however what happens instead is, that the value of EB reduces gradually down to 8\% when field cooling from 15\,K to 10\,K only.

This unusual behavior  reflects the fact that the strength and sign of the magnetic exchange interaction  varies drastically between the parent compounds. E.g. in LaFeO$_3$,  $T_{\mathrm{N}}$ is 740\,K \cite{Eibschutz1967a}; LaCrO$_3$ is 290\,K \cite{Daniels2013,Prado-Gonjal2013}.  In LaMnO$_3$,  antiferromagnetically coupled FM  planes order  below 100\,K \cite{Wollan1955}, while LaNiO$_3$ and LaCoO$_3$ are paramagnetic down to lowest temperatures \cite{Vasanthacharya1984,Yan2004}.   
Therefore the strength of the magnetic exchange and with that the magnetic correlations
can be locally different, depending on the local elemental composition. This would tentatively explain a distribution of magnetic transition temperatures  on a local scale.

 The latter argument on mixed ionic bonds, e.g. magnetic exchange interactions   also sheds some light on the underlying mechanism leading to the VEB behavior. These competing  magnetic exchange interactions will necessarily create magnetic frustration on a very local scale. But in the present case, judging from the $M$($\upmu_0H$) curves,  AFM coupling  still prevails. To support this conclusion, the  magnetic transition was investigated with AC SQUID magnetometry  (see Fig.\,\ref{fig_ex_labo} and the Supplementary material). Such a study  helps to distinguish between  AFM, FM, ferrimagnetic and a possible spin-glass-like frustrated configuration by comparing the frequency dependent magnetic response of the material. This investigation  yields no significant effect of either the  driving frequency or the amplitude of the oscillating field, which thus excludes(see supporting information) (i) a magnetic spin glass state as well as (ii) a purely ferrimagnetic state. Instead, the featureless appearance of $m''$ across the transition points towards a predominant AFM coupling \cite{Baanda2013}. This observation is not in contradiction to the postulated model of small FM clusters embedded in the matrix as they are a minor component and  exchange coupled to the AFM matrix.

 \begin{figure}[t!]
\[\includegraphics[width=0.9\columnwidth]{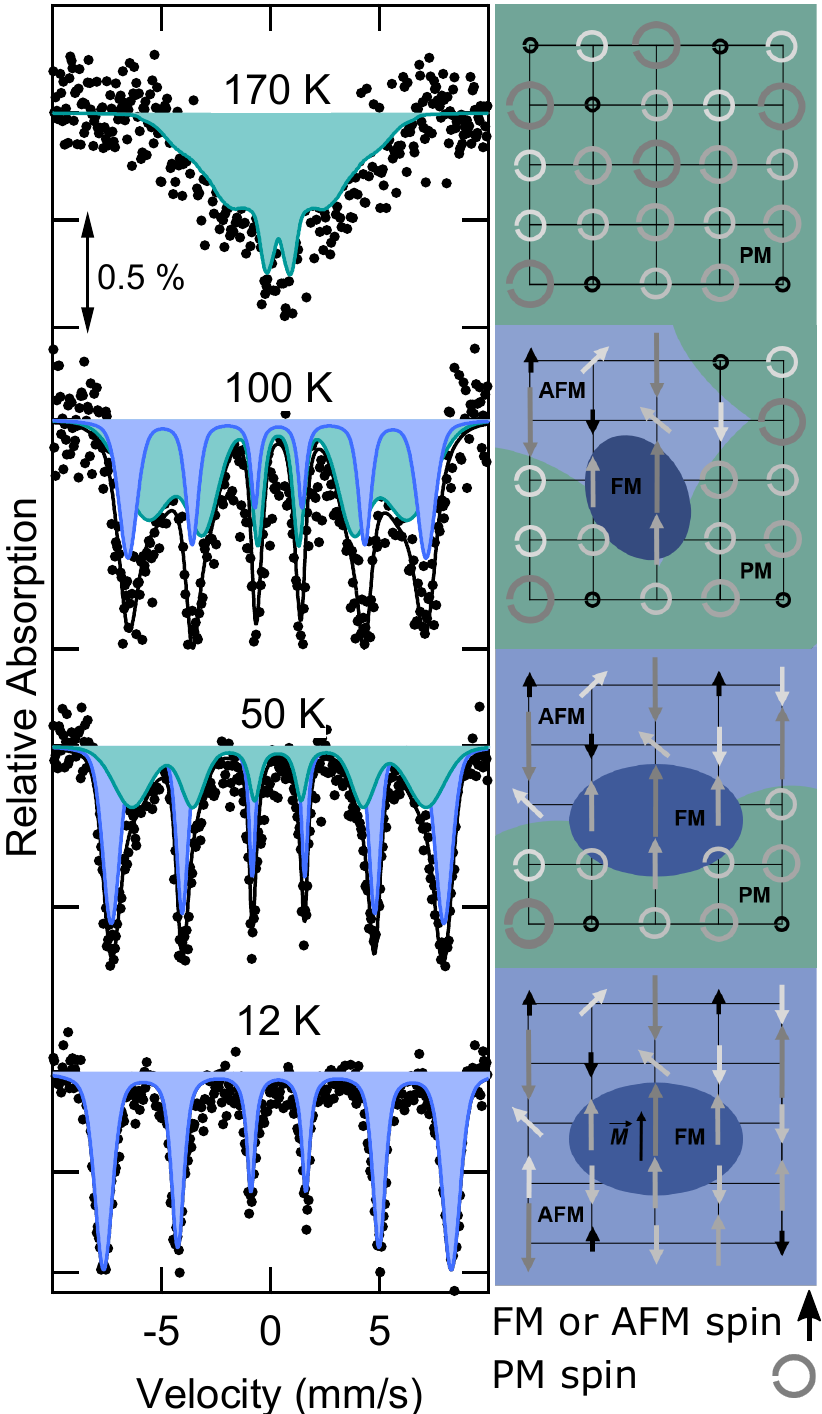}\]
\caption{ Left column: M\"ossbauer spectra in zero magnetic field as function of temperature, represented with two sextets, one  broad spectrum (green) representing dynamic fluctuating spins (on the characteristic timescale of the measurement)  and one well defined subspectrum (blue) from static magnetic order. Right column: A tentative sketch of the evolution of the proposed magnetic structure. At high temperatures spins are dynamically fluctuating (PM, indicated by circles), with decreasing temperature spins start to couple FM and AFM (blue areas), followed by more and more AFM coupling areas. At low temperatures five different kinds of spins are coupled predominantly AF, but one pair of mixed spins couples FM (center). This leads to a local FM cluster which is coupled to the surrounding AFM matrix. Naturally not all exchange interactions can be satisfied, resulting in magnetic frustration and spin canting.}
\label{fig_MS_sketch}
\end{figure}

The magnetometry presented above  already provides  valuable insights into the magnetic properties of PE-HEOs. However, a local element-specific view, as it is possible with $^{57}$Fe M\"ossbauer spectroscopy, will help to underpin the above considerations. Spectra were measured from ambient temperature across the magnetic transition down to 12\,K and selected measurements are shown in Fig.\,\ref{fig_MS_sketch}. At room temperature a quadrupole doublet is observed typical for octahedrally coordinated Fe$^{3+}$ (not shown), while at 12\,K a magnetic splitting is observed with an average hyperfine field $B_{\mathrm{HF}}$ of 49\,T. The considerable value of the average $B_{\mathrm{HF}}$ provides evidence for a large local magnetic moment on the order of several $\upmu_{B}$, which is typical of Fe$^{3+}$.  $^{57}$Fe M\"ossbauer spectroscopy differs from magnetization in that the measured hyperfine field $B_{\mathrm{HF}}$ is independent of direction: the moment orientation with respect to the gamma ray direction only enters into the relative line intensities and not the line separation. 

The broadening of the absorption lines can be well fitted with a Gaussian distribution with a width of 2\,T, representing the chemical disorder around the Fe sites. However the broadening is small compared to other (only) ternary compounds in which Fe has been substituted by e.g. Co, Mn, and/or Cr \cite{Orlinski2017,Jia1994,Kuznetsov2001}, which show  a broad distribution of hyperfine parameters or even separate individual environments. The small broadening clearly shows that the local environment of Fe is surprisingly well-defined, despite the disordered nature of the material. This comparison with ternary compounds directly shows that the HEO approach allows for the stabilization of  single phase materials in the center of complex multi-element phase diagrams, which are otherwise not accessible.

M\"ossbauer spectra measured in a magnetic field of 5\,T parallel to the $\gamma$ beam at 4.3\,K (see spectrum in the Supplementary material), show a partial reorientation of the hyperfine fields towards a perpendicular arrangement with respect to the magnetic field (the area ratio of absorption lines is 3:3:1). Such a behavior is typical of an AFM or canted AFM system. In conclusion the observed small magnetization originates either from a canted AFM arrangement, locally uncompensated spins or small FM clusters or both.

A spectrum measured directly below the magnetic transition temperature at 170\,K clearly shows dynamic relaxation of the magnetic moments on the time scale of the M\"ossbauer experiment (e.g. onset of paramagnetism PM) and is therefore not fitted. A  detailed analysis of the  spectra measured at 50 and 100\,K leads to the conclusion that two subspectra are required to represent the data: One spectrum with large hyperfine splitting and well defined line width, i.e. a similar broadening then observed at 12\,K. The second subspectrum shows large broadening and a collapsing magnetic hyperfine field\footnote{For simplicity we here represent this dynamic subcomponent with Gaussian broadened Lorentzian sextets}. This latter component can be directly attributed to areas of the sample in which magnetic order is dynamic on the timescale of the M\"ossbauer measurement ($\tau \approx 10^{-9}$\,s) at the respective temperature. 
Magnetic relaxation does not alter the subspectrum area, only its shape.
While at 170\,K the entire spectrum is dynamic, the spectral area ratio of the dynamic component decreases from 60 to 40\% when cooling from 100\,K to 50\,K, while only one well defined sextet is sufficient to represent the data at 12\,K.

Summarizing the results on the purely TM based magnetism in  La(5B$_{0.2}$)O$_3$ one can state that below 185\,K magnetic ordering sets in gradually. The complex magnetic state is also responsible for the peculiar occurrence of VEB in this structurally single phase system.    In order to observe VEB it is necessary that small ferromagnetically coupling clusters start to order at higher temperatures and are locked into the gradually ordering AFM matrix. The proposed evolution of the magnetic structure is illustrated in the sketch in Fig.\,\ref{fig_MS_sketch}.

This intricate magnetic behavior makes the system interesting and unique. Further studies employing wide and small angle neutron scattering for identfying local fluctuations of the magnetization and the magnetic structure, X-ray magnetic dichroism for the element specific temperature  evolution of the magnetic moments, and possibly also local nuclear spectroscopic techniques, such as nuclear forward scattering,  may help in describing in full the magnetic structure of this complex system.

\subsection{A(Co$_{0.2}$Cr$_{0.2}$Fe$_{0.2}$Mn$_{0.2}$Ni$_{0.2}$)O$_3$}
\label{sec:A_5B0,2_O3}

\begin{figure}[htb]
\[\includegraphics[width=0.7\columnwidth]{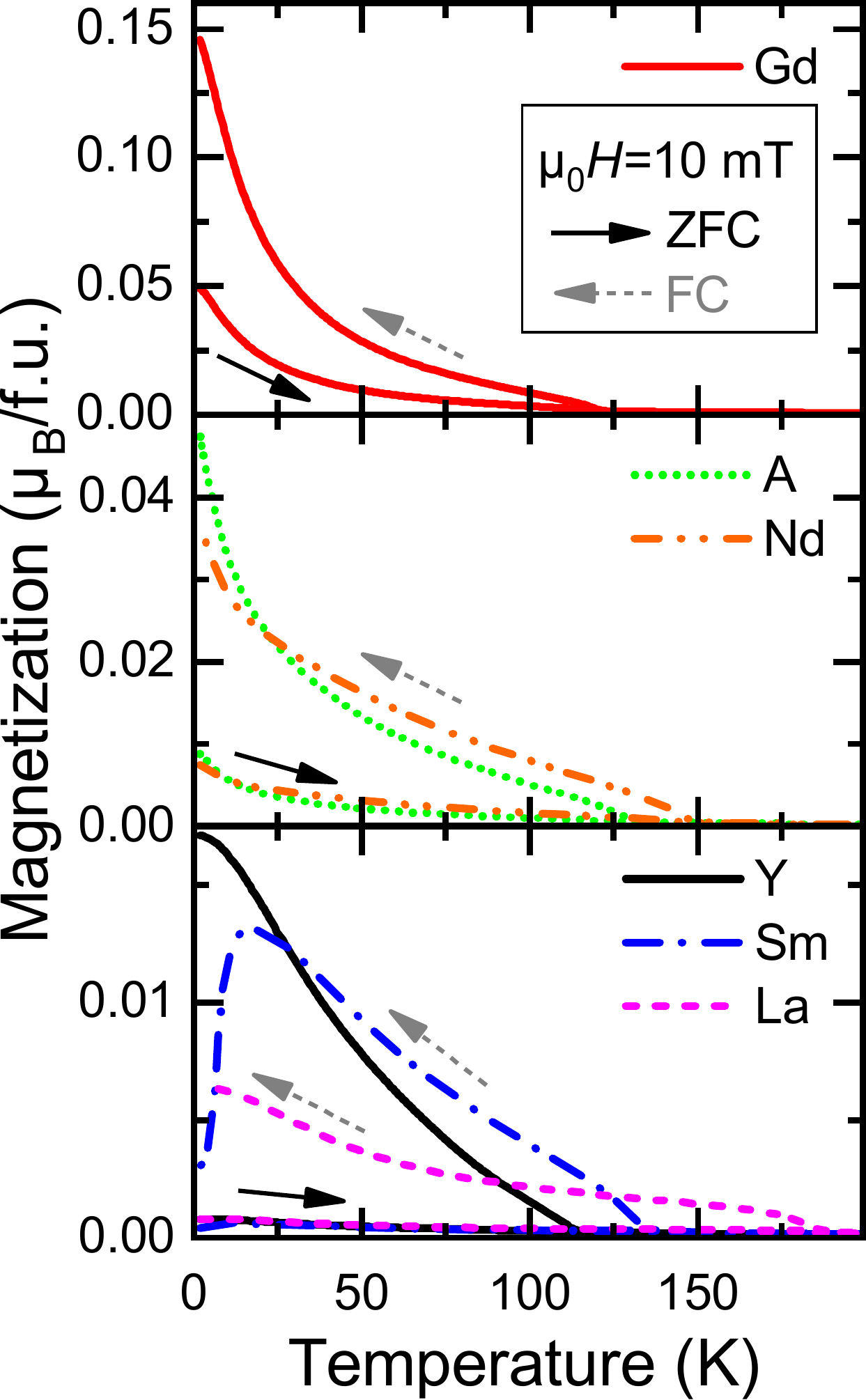}\]
\caption{Magnetization of A(5B$_{0.2}$)O$_3$ as a function of temperature, measured after zero field cooling and field cooling in $\upmu_0H$=10\,mT. The curves are grouped with respect to their magnitude.}
\label{fig_m_amn2}
\end{figure}

\begin{figure}[htb]
\[\includegraphics[width=0.8\columnwidth]{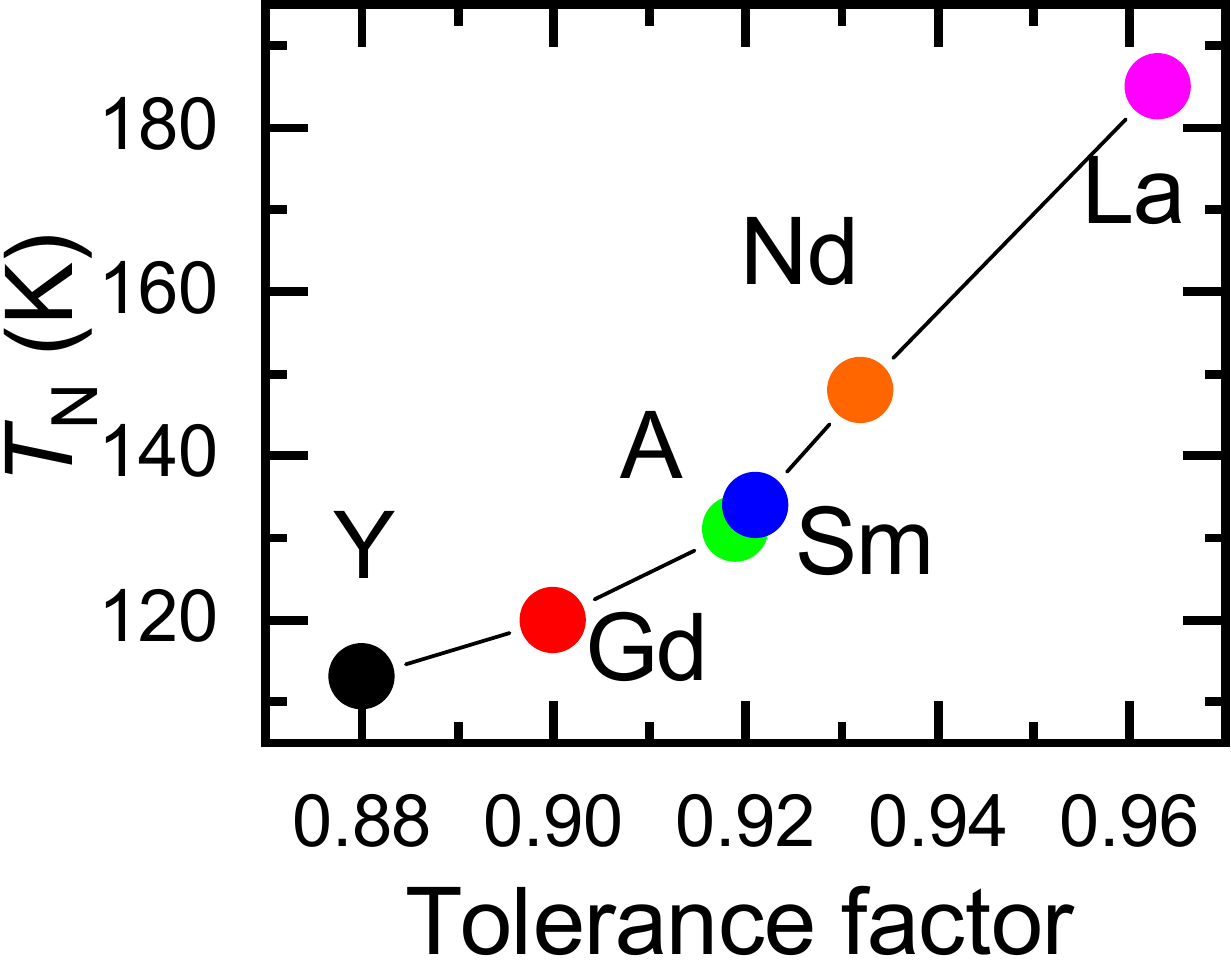}\]
\caption{Magnetic transition temperature  of the samples as function of the tolerance factor of the structures and the RE element. The line serves as guide to the eye.}
\label{fig_tolerance}
\end{figure}

Samples A(Co$_{0.2}$Cr$_{0.2}$Fe$_{0.2}$Mn$_{0.2}$Ni$_{0.2}$)O$_3$  with A=Gd, La, Nd, Sm, Y or  5A$_{0.2}$ have been characterized by magnetometry. Temperature dependent ZFC and FC curves are presented in Fig.\,\ref{fig_m_amn2}. These have been grouped according to their magnitude. The  magnetic transition temperatures (here denoted as N\'eel temperatures $T_{\mathrm N}$) are plotted as a function of the Goldschmidt tolerance factor in Fig.\,\ref{fig_tolerance}.
All the $M$($T$) curves clearly indicate magnetic ordering transitions, which are in the temperature range from 110 to 180\,K. All samples, with exception of the Sm compound, show a continuous increase in  $M$($T$)  with decreasing temperature. In the compounds with magnetic  REs ions (Gd, Nd, 5A$_{0.2}$), their large magnetic moment becomes visible at low temperatures. The Sm(5B$_{0.2}$)O$_3$ compound however, shows a decrease of the magnetization towards low temperatures, leading nearly to a magnetization reversal. A similar behavior has also been observed in SmFeO$_{3}$ \cite{Lee2011a} and has been  attributed to long-range ordering of Sm$^{3+}$ spins, which couple antiferromagnetically to the canted magnetic moment of the Fe$^{3+}$ ion. A drop in the magnetization is also observed in orthochromites at low temperatures, which in this case has been related to a spin-reorientation transition of the antiferromagnetically coupled Cr spins \cite{Rajeswaran2012a}. What exactly is happening in case of the Sm(5B$_{0.2}$)O$_{3}$ compound is not straightforwardly deducible from general principles. In order to get a more detailed physical picture,  the spin structures need to be fully resolved and element specific magnetic moments deduced.

What can  also be noticed when comparing the two compounds with no magnetic moment on the RE site, namely, Y and La, is that the difference between FC and ZFC is much larger for the Y compound, also the magnetization reached after FC is a factor of three times larger. This is again an interesting finding as it illustrates the importance of the structural features for the magnetic properties, as the RE magnetism plays no role. The first observation can be interpreted as an indication of larger magnetic anisotropy in the structurally more distorted Y compound, which might be a reasonable explanation since locally the TM octahedra are strongly anisotropic themselves. This might in turns result in a locally large anisotropy. The underlying reason for the second observation remains unclear as it is directly linked to the open question about the origin of the observed net magnetization, whether stemming from a frustrated system, locally uncompensated ferrimagnetic spins or small ferromagnetic clusters.

An interesting finding is that the magnetic transition temperatures for the six compounds are directly correlated to their Goldschmidts tolerance factor (see Fig.\,\ref{fig_tolerance}). This factor is a  measure of the distortion of the crystal lattice for different RE ions and is strongly related to the B-O-B bond angle,  decreasing with decreasing tolerance factor. A direct correlation of magnetic transition temperatures to structural and electronic characteristics has also been found in the RE-orthoferrites \cite{Treves1965a,Lyubutin1999}, RE-orthochromites \cite{Zhou2010a} and RE-orthomanganites \cite{Zhou2006}. However a simple geometric  relation to the bond angles is only found in the ferrites  while in the chromites and maganites  also other orbital overlap integrals need to be considered due to the different electron configuration of the cations\cite{Zhou2010a}. Thus it is  interesting that also in the chemically disordered lattice of the high entropy oxides, in which ions with different electronic configuration magnetically interact, such a continuous structural dependency is observed.

\section{Conclusion}\label{sec:sum}

The first comprehensive study on the magnetic properties of RE and TM based perovskite type high entropy oxides is presented. 
Compounds with intermixed B-site (five TM elements) and single element A-sites (RE elements) as well as a compound with five different RE elements on the A-sites were investigated (decenary compound). It was found that the magnetic properties of these novel compounds \cite{Sarkar2017a} can only be explained by the presence of competing magnetic exchange interactions within the TM cation sublattice. Detailed investigations by magnetometry and element specific M\"ossbauer spectroscopy evidence a complex magnetic state, which is mainly dominated by AF interactions. However a large degree of magnetic frustration  is found due to the high degree of disorder and competing FM and AF interactions. It is proposed that the sign of the magnetic exchange interactions locally alternates, leading to  small FM clusters within the predominantly AF matrix. This nanoscale magnetic feature may be responsible for the vertical exchange bias of about 25\% of the remnant magnetization.
 Considering the chemical disorder, we find  it  surprising that the magnetic ordering temperature of the compounds is directly controlled by the size of the RE ion (Goldschmidt tolerance factor). This correlation 
can be utilized to 
provide  a method for fine adjusting of the magnetic transition temperatures in these compounds.

The  concept of high entropy  multi-element oxides  allows for stabilization of compounds and structures beyond the doping regime,  not accessible otherwise and which 
can 
feature unprecedented novel properties. It 
is 
anticipated that also other physical properties, such as dielectric or magneto-transport properties will be tailored with great freedom.
Understanding the underlying mechanisms of the properties of these multi-element materials will need joint efforts of experimental and theoretical scientists.  The vast multidimensional research space will require the use of experimental and theoretical high throughput methods, such as adapted combinatoric synthesis methods \cite{Koinuma2004,Amis2004} and high throughput ab-initio calculations employing the appropriate choice of descriptors \cite{Curtarolo2013} in order to identify promising candidates for applications. Many other outstanding physical properties are expected considering that the class of parent compounds is known for their spectacular properties, among them ferroelectricity and giant magnetocapacitance \cite{Goto2004}, multiferroic order \cite{Cheong2007}, colossal magneto resistance \cite{Ramirez1997a} or magnetocaloric \cite{Das2017} properties.

\begin{acknowledgments}
  We acknowledge funding by financial support from the Helmholtz Association and the Deutsche Forschungsgemeinschaft (DFG) project HA 1344/43-1 and WE 2623/14-1. We acknowledge discussion with O. Clemens (TU Darmstadt). 

\end{acknowledgments}


%

\end{document}